\documentclass[twocolumn,english,reprint,showpacs,preprintnumbers,amsmath,amssymb,aps]{revtex4}
\usepackage[T1]{fontenc}
\usepackage[latin9]{inputenc}
\usepackage[active]{srcltx}
\usepackage{xcolor}
\usepackage{textcomp}
\usepackage{amsmath}
\usepackage{amssymb}
\usepackage{esint}
\PassOptionsToPackage{normalem}{ulem}
\usepackage{ulem}

\makeatletter

\providecolor{lyxadded}{rgb}{0,0,1}
\providecolor{lyxdeleted}{rgb}{1,0,0}

\@ifundefined{textcolor}{}
{%
 \definecolor{BLACK}{gray}{0}
 \definecolor{WHITE}{gray}{1}
 \definecolor{RED}{rgb}{1,0,0}
 \definecolor{GREEN}{rgb}{0,1,0}
 \definecolor{BLUE}{rgb}{0,0,1}
 \definecolor{CYAN}{cmyk}{1,0,0,0}
 \definecolor{MAGENTA}{cmyk}{0,1,0,0}
 \definecolor{YELLOW}{cmyk}{0,0,1,0}
}

\usepackage{dcolumn}
\usepackage{bm}
\usepackage{amsfonts}\usepackage{babel}
\providecommand{\U}[1]{\protect\rule{.1in}{.1in}}

\@ifundefined{textcolor}{}{
\definecolor{BLACK}{gray}{0}
 \definecolor{WHITE}{gray}{1}
 \definecolor{RED}{rgb}{1,0,0}
 \definecolor{GREEN}{rgb}{0,1,0}
 \definecolor{BLUE}{rgb}{0,0,1}
 \definecolor{CYAN}{cmyk}{1,0,0,0}
 \definecolor{MAGENTA}{cmyk}{0,1,0,0}
 \definecolor{YELLOW}{cmyk}{0,0,1,0}
 }

\usepackage{babel}
\usepackage{ragged2e}

\usepackage{babel}

\makeatother

\usepackage{babel}
\begin{document}

\title{Space-filling branes of gravitational ancestry}

\author{Claudio Bunster}

\email{bunster@cecs.cl}

\affiliation{Centro de Estudios Científicos (CECs), Avenida Arturo Prat 514, Valdivia,
Chile}

\author{Alfredo Pérez}

\email{aperez@cecs.cl}

\affiliation{Centro de Estudios Científicos (CECs), Avenida Arturo Prat 514, Valdivia,
Chile}

\begin{abstract}
We introduce a new kind of space-filling brane, which we term ``G-brane''
because its action is a descendant of the gravitational action. The
G-brane may be thought of as the remanent of the gravitational field
when the propagating gravitons are removed. The G-brane is different
from the Dirac or Nambu space-filling branes. Its properties in any
spacetime dimension {\normalsize{}$D$} are exhibited. When the spacetime
dimension $D$ is greater than or equal to three, the G-brane does
not possess propagating degrees of freedom, just as the Dirac or Nambu
branes. For $D=3$ the G-brane yields a reformulation of gravitation
theory in which the Hamiltonian constraints can be solved explicitly,
while keeping the spacetime structure manifest. For $D=2$ the G-brane
provides a realization of the conformal algebra, i.e. a conformal
field theory, in terms of two scalar fields and their conjugates,
which possesses a classical central charge. In the G-brane reformulation
of (2+1) gravity, the boundary degrees of freedom of the gravitational
field in asymptotically anti--de Sitter space appear as ``matter''
coupled to the (1+1) G-brane on the boundary.
\end{abstract}
\maketitle

\section{Introduction}

The existence itself of the black hole, and the ensuing mystery of
the origin of the black hole entropy, show that there is significant
gravitational physics in vacuum which is not due to propagating gravitons.
This verity is strikingly evidenced by the existence of a black hole
in 2+1 spacetime dimensions \cite{BTZ,HBTZ}. One would like therefore
to have a theory which captures what is left of gravitation after
the propagating gravitons are removed. To be of general applicability
such a descendant of general relativity should retain the general
covariance of its ancestor.

We propose herein that the ``gravitational field without propagating
gravitons'' may be described by a new kind of space-filling brane,
which we term ``G-brane,'' whose action is a direct descendant of
the action for the gravitational field. The G-brane is different from
the Dirac or Nambu space-filling branes.

We develop the general G-brane formalism for an arbitrary spacetime
dimension $D$ and focus in detail on the lower dimensions $D=3$
and $D=2$. For $D=3$ the G-brane yields a reformulation of gravitation
theory in which the Hamiltonian constraints can be solved explicitly,
while keeping the spacetime structure manifest. For $D=2$, the G-brane
provides a realization of the conformal algebra, i.e. a conformal
field theory, in terms of two scalar fields and their conjugates,
which possesses a classical central charge. In the G-brane reformulation
of (2+1) gravity, the boundary degrees of freedom of the gravitational
field in asymptotically anti--de Sitter space appear as ``matter''
coupled to the (1+1) G-brane on the boundary.

Many years ago \cite{Regge-Teitelboim-strings} an attempt was made
to reformulate Einstein's theory of gravitation in $D$ dimensions,
as the theory of a \emph{non-space-filling} brane, embedded in a higher
dimensional fixed (flat) spacetime background. However, obstacles
that could not be satisfactorily circumvented were met (see also \cite{Deser-Pirani-Robinson}
and, for later developments, \cite{Paston-Semenova1,Paston-semenova2}).
Those difficulties are not present for the space-filling brane.

The plan of the paper is the following. Section \ref{sec:Dirac-and-Nambu}
reviews and puts in context the concepts of Dirac and Nambu space-filling
branes. Section \ref{sec:G-brane} introduces the G-brane, and discusses
its general properties for $D\geq3$, while Sec. \ref{sec:G-brane-for}
discusses the special case of the G-brane for $D=2$. Section \ref{sec:Gravitation-in-three}
contains the reformulation of gravitation in three spacetime dimensions
as the theory of a G-brane. Finally Sec. \ref{sec:Concluding-remarks}
is devoted to conclusions and possible further developments along
this line of inquiry.

\section{Dirac and Nambu space-filling branes\label{sec:Dirac-and-Nambu}}

\subsection{Surface deformations and surface variables }

Space-filling branes were first introduced by Dirac in his formulation
of field dynamics on an arbitrary spacelike surface, within a fixed
background spacetime that he mostly took to be flat \cite{Dirac-1}.
He introduced \emph{surface variables} $y^{\lambda}\left(x,t\right)$,
which describe the embedding of a parametrized spacelike surface with
respect to a fixed system of coordinates $y^{\lambda}$ on the background,
and their conjugate momenta $w_{\lambda}$. Thus, in terms of Poisson
brackets, 
\[
\left[y^{\lambda}\left(x\right),w_{\text{\ensuremath{\rho}}}\left(x'\right)\right]=\delta_{\rho}^{\lambda}\delta\left(x,x'\right).
\]

The Hamiltonian for a field theory in which the $t=\text{constant}$
surface is a generic spacelike surface was then written as
\[
H=\int dxN^{\lambda}h_{\lambda},
\]
where the generators $h_{\lambda}$ have the form 
\begin{equation}
h_{\lambda}=w_{\lambda}-T_{\lambda\rho}n^{\rho}.\label{eq:hlambdaDirac}
\end{equation}

The action reads

\[
I=\int dxdt\left(w_{\lambda}\dot{y}^{\lambda}+\pi\dot{\varphi}-N^{\lambda}h_{\lambda}\right).
\]

Here $n^{\rho}$ is the unit normal to the surface,
\begin{align}
\gamma_{\lambda\rho}n^{\lambda}n^{\rho} & =-1,\label{eq:n1}\\
n_{\lambda}y_{\;\;,i}^{\lambda} & =0.\label{eq:n2}
\end{align}

The components $n^{\rho}$ are functions of the $y$'s and their first
spatial derivatives. If $n^{\rho}$ solves (\ref{eq:n1}) and (\ref{eq:n2}),
so does $-n^{\rho}$ corresponding to time reversal. We have denoted
by $\gamma_{\lambda\rho}$ the metric in the external coordinate system
in order to reserve the letter $g$ for the spatial metric $g_{ij}$
induced on the parametrized surface:
\begin{equation}
g_{ij}\left[y\right]\left(x\right)=\gamma_{\lambda\rho}\left(y\left(x\right)\right)y_{\;\;,i}^{\lambda}y_{\;\;,j}^{\rho}.\label{eq:saptialmetric}
\end{equation}
The $T_{\lambda\rho}$ are the components of the symmetric energy-momentum
tensor of the matter fields at hand, with appropriate density weight.
The projections $T_{\perp\perp}=T_{\lambda\rho}n^{\lambda}n^{\rho}$
and $T_{\perp i}=T_{\lambda\rho}n^{\lambda}y_{\;\;,i}^{\rho}$, which
are the energy-momentum densities, depend on the matter canonical
variables $\pi,\varphi$, and on the $y^{\lambda}$, but they do not
depend on the $w_{\lambda}$.

The $N^{\lambda}$ are Lagrange multipliers for the constraints,

\[
h_{\lambda}\approx0,
\]
in Dirac's ``weakly vanishing'' terminology and notation \cite{Dirac-2}.

The generators $h_{\lambda}$ may be decomposed into a component $h_{\perp}$
normal to the surface and $h_{i}$ tangential to it, that is: 
\begin{align}
h_{\perp} & =\left(w_{\lambda}-T_{\lambda\rho}n^{\rho}\right)n^{\lambda}=w_{\perp}-T_{\perp\perp},\label{eq:hperpDirac}\\
h_{i} & =\left(w_{\lambda}-T_{\lambda\rho}n^{\rho}\right)y_{\;\;,i}^{\lambda}=w_{i}-T_{\perp i}.\label{eq:hiDirac}
\end{align}

The corresponding multipliers are then the ``lapse,'' and ``shift''
functions, which are related to the $N^{\lambda}$ through

\[
N^{\lambda}=N^{\perp}n^{\lambda}+N^{i}y_{\;\;,i}^{\lambda}.
\]

The equations of motion of the $y^{\lambda}$ are
\[
\dot{y}^{\lambda}=N^{\perp}n^{\lambda}+N^{i}y_{\;\;,i}^{\lambda},
\]
which, since $N^{\perp}$ and $N^{i}$ are arbitrary functions of
time, express the fact that any deformation of the brane is permissible.
The equations of motion for the matter fields describe their response
to a surface deformation.

\subsection{Surface deformation algebra}

The perpendicular and tangential projections are useful because $h_{\perp}$
generates deformations of the surface which are of dynamical importance,
whereas $h_{\perp i}$ merely reparametrizes the spacelike surface
without really moving away from it. Note that this decomposition makes
essential use of the spacetime metric in order to give an unambiguous
meaning, through the normal, to the notion of ``moving away from
the surface.'' Thus the notion of surface deformation expressed in
terms of the projected generators is different from, and has much
more structure than, the notion of a spacetime diffeomorphism. The
surface deformation generators must obey integrability conditions
that guarantee that the evolution is ``path independent'': if one
starts from a given initial spacelike surface and ends on a given
final spacelike surface, the result should be independent of the intermediate
sequence of surfaces employed to achieve the total deformation as
a sequence of infinitesimal ones \cite{Teitelboim1,Teitelboim2}.
These integrability conditions read, in terms of the unprojected generators,
\begin{equation}
\left[h_{\lambda}\left(x\right),h_{\rho}\left(x'\right)\right]=0,\label{eq:habelian}
\end{equation}
or in terms of the projected ones, 
\begin{align}
\left[h_{\perp}\left(x\right),h_{\perp}\left(x'\right)\right] & =g^{ij}\left(h_{i}\left(x\right)+h_{i}\left(x'\right)\right)\delta_{,j}\left(x,x'\right),\label{eq:hphp}\\
\left[h_{i}\left(x\right),h_{\perp}\left(x'\right)\right] & =h_{\perp}\left(x\right)\delta_{,i}\left(x,x'\right),\label{eq:hphi}\\
\left[h_{i}\left(x\right),h_{j}\left(x'\right)\right] & =h_{i}\left(x'\right)\delta_{,j}\left(x,x'\right)+h_{i}\left(x\right)\delta_{,i}\left(x,x'\right).\label{eq:hihi}
\end{align}
The last two equations, which involve $h_{i}$ on the left-hand side
are of purely kinematical nature since they just capture how $h_{\perp}$
and $h_{i}$ behave under changes in the spatial coordinates. The
truly restrictive equation is the first one (\ref{eq:hphp}) which
imposes severe conditions on the components of the energy-momentum
densities. Its quantum mechanical counterpart was referred to by Schwinger
in the concluding sentence of \cite{Schwinger} as\emph{ \textquotedblleft what
may well be considered the most fundamental equation of relativistic
quantum field theory.\textquotedblright{}}

\subsection{Dirac brane}

If one drops the matter fields in the above equations, one has
\[
T_{\lambda\rho}=0,
\]
and the generators read
\[
h_{\lambda}=w_{\lambda},
\]
or in their projected version,
\begin{align*}
h_{\perp}= & w_{\perp},\\
h_{i}= & w_{i}.
\end{align*}
The generators $w_{\perp}$ and $w_{i}$ obey the algebra (\ref{eq:hphp})--(\ref{eq:hihi})
for any background, and they generate a consistent evolution for a
space-filling brane that we will call a \emph{Dirac brane}. Equations
(\ref{eq:hlambdaDirac}), (\ref{eq:hperpDirac}) and (\ref{eq:hiDirac})
may then be said to describe the coupling of a Dirac brane to matter.

For the Dirac brane without matter fields, the action in Hamiltonian
form reads 
\[
I=\int dxdt\left(w_{\lambda}\dot{y}^{\lambda}-N^{\lambda}w_{\lambda}\right),
\]
where $g$ is the determinant of the spatial metric $g_{ij}$. If
one passes to the Lagrangian form of the action by eliminating $w_{\lambda}$
from the constraint, one gets
\[
I_{\text{}}=0.
\]

\subsection{Nambu brane}

A different kind of space-filling brane, which we shall call \emph{Nambu
brane} for reasons that will become evident in a moment, is obtained
by demanding that there should be a constant energy density $\mathcal{N}$
through space, that is, one sets, 
\[
T_{\lambda\rho}=\mathcal{N}g^{1/2}\gamma_{\lambda\rho}.
\]

The generators now read
\begin{equation}
h_{\lambda}=w_{\lambda}-\mathcal{N}g^{1/2}n_{\lambda},\label{eq:cnambu}
\end{equation}
or
\begin{align*}
h_{\perp} & =w_{\perp}+\mathcal{N}g^{1/2},\\
h_{i} & =w_{i},
\end{align*}
and they obey the integrability conditions (\ref{eq:habelian})--(\ref{eq:hihi}),
for any background.

For the Nambu brane the action in Hamiltonian form may be written
as 
\begin{align*}
I & =\int dxdt\left(w_{\lambda}\dot{y}^{\lambda}-N^{\lambda}\left(w_{\lambda}-\mathcal{N}g^{1/2}n_{\lambda}\right)\right),
\end{align*}
whereas the Lagrangian form now reads
\begin{align}
I & =\mathcal{N}\int dxdt\left(g^{1/2}n_{\lambda}\dot{y}^{\lambda}\right),\nonumber \\
 & =-\mathcal{N}\int dxdt\left(-g_{\text{spacetime}}\right)^{1/2},\label{eq:nambu}
\end{align}
which is indeed the action for a (space-filling) Nambu brane. If one
has a matter energy-momentum tensor, obeying the appropriate Poisson
bracket rules which make the Dirac brane generators (\ref{eq:hperpDirac})
and (\ref{eq:hiDirac}) to close according to (\ref{eq:hphp})--(\ref{eq:hihi}),
then its addition to the Nambu brane generators will preserve the
closure.

The Lagrangian action for the Nambu brane (\ref{eq:nambu}) is invariant
under spacetime reparametrizations, 
\[
y^{\lambda}\left(x,t\right)\rightarrow y^{\lambda}\left(x,t\right)+\delta y^{\lambda}\left(x,t\right).
\]
More precisely, if the integral in (\ref{eq:nambu}) is extended over
a spacetime region $M$ with boundary $\partial M$ one has
\begin{equation}
\delta I_{\text{}}=\int_{\partial M}dx\left(\mathcal{N}g^{1/2}n_{\lambda}\delta y^{\lambda}\right),\label{eq:deltInambu}
\end{equation}
which is where (\ref{eq:cnambu}) may be thought to come from in the
first place.

We now turn our attention to the main point of the paper, a new type
of brane.

\section{G-brane for $D\geq3$\label{sec:G-brane}}

\subsection{Action}

A different type of space-filling brane is obtained by replacing the
spacetime volume (\ref{eq:nambu}) by a different invariant, the action
for the gravitational field evaluated on a given background. Hence
the name G-brane.

The Lagrangian action for the gravitational field is
\begin{equation}
I_{\text{grav}}=\int dtdxN^{\perp}\mathcal{L},\label{eq:ancestor}
\end{equation}
with 
\begin{align}
\mathcal{L} & =\frac{1}{2\kappa}\left(G^{ijkl}K_{ij}K_{kl}+g^{1/2}\left(R-2\Lambda\right)\right),\label{eq:L}
\end{align}
and
\[
G^{ijkl}=\frac{1}{2}g^{1/2}\left(g^{ik}g^{jl}+g^{il}g^{jk}-2g^{ij}g^{kl}\right).
\]
Here $K_{ij}$ denotes the extrinsic curvature of the $t=\text{constant}$
surface, which is the ``invariant velocity of $g_{ij}$''
\[
\dot{g}_{ij}=-2N^{\perp}K_{ij}+N_{i/j}+N_{j/i},
\]
and $\kappa$ is related to Newton's gravitational constant by $\kappa=8\pi G$. 

The action (\ref{eq:L}) differs with the Hilbert action by adding
to it a divergence which removes the second time derivatives of $g_{ij}$
and also the second and first derivative of $g_{0\mu}$. 

Under a spacetime reparametrization the analog of (\ref{eq:deltInambu})
is
\begin{align}
\delta I & =-\int_{\partial M}dx\left[\mathcal{L}n_{\lambda}\delta y^{\lambda}\right],\label{eq:IGbrane-1}
\end{align}
which yields the G-brane generators,
\begin{equation}
h_{\lambda}=w_{\lambda}+\mathcal{L}n_{\lambda},\label{eq:hlambda}
\end{equation}
whose projected version is
\begin{align}
h_{\perp} & =w_{\perp}-\mathcal{L},\label{eq:hpRT}\\
h_{i} & =w_{i}.\label{eq:HIRT2}
\end{align}
Here the extrinsic curvature is understood to be expressed in terms
of the surface variables $y^{\lambda}$, by the counterpart, 
\begin{equation}
K_{ij}=n_{\lambda}D_{i}y_{\;\;,j}^{\lambda},\label{eq:extrinsic curv}
\end{equation}
of (\ref{eq:saptialmetric}), where $D_{j}=y_{\;\;,j}^{\lambda}D_{\lambda}$
is the covariant derivative in the external space projected on the
$t=\text{constant}$ surface.

For the G-brane the action in Hamiltonian form reads 
\begin{align*}
I & =\int dxdt\left[w_{\lambda}\dot{y}^{\lambda}-N^{\lambda}\left(w_{\lambda}+\mathcal{L}n_{\lambda}\right)\right],
\end{align*}
and the Lagrangian form is

\begin{align}
I_{\text{}} & =-\int dxdt\left[\mathcal{L}n_{\lambda}\dot{y}^{\lambda}\right].\label{eq:IGbrane}
\end{align}

Equations (\ref{eq:hpRT}) and (\ref{eq:HIRT2}) were obtained many
years ago \cite{Regge-Teitelboim-strings} in an attempt to reformulate
Einstein's gravity in $D$ dimensions as the theory of a \emph{non-space-filling}
brane embedded in a higher dimensional\emph{ }(flat) spacetime background.
In that case one cannot express the extrinsic curvature in terms of
the surface variables through (\ref{eq:extrinsic curv}) and the need
for imposing additional constraints arises. It was proposed in \cite{Regge-Teitelboim-strings}
that these additional constraints might be the standard gravitational
initial valued constraints, but the analysis was not carried to completion
(see also \cite{Deser-Pirani-Robinson} and, for later developments,
\cite{Paston-Semenova1,Paston-semenova2}).

\subsection{Integrability and closure}

The generators $h_{\lambda}$ given by (\ref{eq:hlambda}) close according
to (\ref{eq:habelian}), which is equivalent to (\ref{eq:hphp})--(\ref{eq:hihi})
because on account of (\ref{eq:IGbrane-1}) one has

\begin{equation}
\mathcal{L}n_{\lambda}=-\frac{\delta I}{\delta y^{\lambda}\left(x\right)}.\label{eq:Ln}
\end{equation}
Here $I$ is the gravitational action evaluated on the given background,
for a spacetime region bounded by an initial and a final spacelike
surface, regarded as a functional of the location of the final surface
$y^{\lambda}\left(x\right)$.

It is essential for the action integral $I$ to be well defined as
a functional of the final surface, that it be invariant under reparametrizations
of the spacetime region in between the two surfaces. This invariance
guarantees that the value obtained for the action is independent of
the particular foliation by $t=\text{constant}$ surfaces employed
in carrying out the integration. 

Note that in order for the resulting G-brane action to depend only
on the $y^{\lambda}$ and his first time derivatives, the associated
gravitational background action should depend only on the $g_{ij}$
and his first time derivatives. 

For $D=3$ and $D=4$ spacetime dimensions the only possibility for
the action $I$ appearing in (\ref{eq:Ln}) is (\ref{eq:L}), which
may be understood as the ``dimensional continuation'' of the Euler
classes of the lower even dimensions $D=0$, corresponding to the
cosmological term (Nambu brane) and $D=2$ (gravitational action without
cosmological term).

For higher spacetime dimensions $D$ there are more possible background
actions, the so-called Lovelock Lagrangians \cite{Lovelock}, which
may be understood as the dimensional continuation of all the Euler
classes of the even dimensions below $D$. The G-brane action arising
from the generic Lovelock Lagrangian is then a sum of terms each of
which involves powers of the intrinsic Riemann tensor of the surface
and its extrinsic curvature \cite{TZ-Lovelock}.

\subsection{Time reversal}

The ancestor action (\ref{eq:ancestor}) is invariant under time reversal
$t\rightarrow-t$ because, although $K_{ij}\rightarrow-K_{ij}$ the
Lagrangian (\ref{eq:L}) is quadratic in $K_{ij}$ ($g_{ij}$ and
$N^{\perp}$ are invariant). However the generator $h_{\perp}$ given
by (\ref{eq:hpRT}) is not invariant because under time reversal $n^{\lambda}\rightarrow-n^{\lambda}$
and therefore $w_{\perp}$ changes sign while $\mathcal{L}$ does
not. The situation is exactly the same as the one for a relativistic
particle where 
\[
p_{\lambda}p^{\lambda}+m^{2}=0
\]
is equivalent to 
\[
p^{0}\mp\sqrt{\vec{p}^{2}+m^{2}}=0.
\]
Here as well the G-brane generators have two branches, which are automatically
included in the description, because Eq. (\ref{eq:hlambda}) does
not specify which of the two solutions of the equations (\ref{eq:n1})
and (\ref{eq:n2}) differing by a sign is chosen.

One may replace the generator (\ref{eq:hpRT}) by the combination
\begin{equation}
h=w_{\lambda}w^{\lambda}+\mathcal{L}^{2},\label{eq:quadraticgen}
\end{equation}
an expression whose appearance is familiar from the bosonic string.
The quadratic generator (\ref{eq:quadraticgen}) incorporates both,
$h_{\perp}$ and its time reversed version. There are cases of interest
for which quadratic the form (\ref{eq:quadraticgen}) is considerably
simpler than its linear counterpart, notably the Born-Infeld electrodynamics
as reformulated by Dirac \cite{DBI}. However from a geometrical point
of view (\ref{eq:quadraticgen}) has the drawback that the bracket
of two such generators at different spatial points close in a way,
which is not universal (the structure coefficients of the constraint
algebra depend on the matter fields).

\subsection{Background energy}

Formally the G-brane, and the Nambu brane, are related to the Dirac
brane by a point canonical transformation, whose generating functional
is precisely $I$ appearing in Eq. (\ref{eq:Ln}).

However, already for the case of the Nambu brane, one has reason to
believe that if the framework is enlarged a bit, the vacuum energy
represented by the cosmological term cannot simply be subtracted.
One knows, for example, that if one introduces $(D-2)$ branes endowed
with a $U\left(1\right)$ charge, the reservoir of vacuum energy can
be tapped \cite{BT1,BT2}, and it even can be used to form black holes
\cite{TGHW}.

Although at present no similar mechanism is known for the other terms
in $\mathcal{L}$, they also represent a sort of ``gravitational
energy of the background,'' which prudence would advise to retain. 

Of course the value of the terms in $g^{-1/2}\mathcal{L}$ other than
the cosmological term $-\kappa^{-1}\Lambda$ depend on the surface
$y^{\lambda}\left(x\right)$. In flat spacetime one may choose a surface
of constant Minkowskian time and both the intrinsic and extrinsic
curvatures will then vanish. However, if one is on a curved background,
a surface where $\mathcal{L}$ vanishes will in general not exist.
For example, in an expanding section of the de Sitter universe, the
intrinsic curvature term vanishes, but the extrinsic one does not.
It, rather, doubles the cosmological term contribution.

Conceptually, the issue of the background energy is not different
from the discussion of the dependence of the inertia of a body upon
its energy content \cite{Einstein}. In that case, if one stays within
the strict context of the dynamics of a relativistic particle that
does not decay, whether one favors the more elegant expression,

\[
E=\frac{m_{0}}{\sqrt{1-v^{2}/c^{2}}}
\]
over
\[
E=\frac{m_{0}}{\sqrt{1-v^{2}/c^{2}}}-m_{0}c^{2},
\]
which reduces to the nonrelativistic expression when $c\rightarrow\infty$,
is a matter of taste, with no physical implication.

But if the particle can decay, so that the rest mass reservoir can
be tapped, then the first choice becomes mandatory.

Before leaving this point we remark that $\mathcal{L}n_{\lambda}$
cannot be interpreted as the components $T_{\perp\lambda}$ of a tensor.
In particular they bear no relation to the conserved energy-momentum
tensor $T^{\lambda\rho}$ of the G-brane, obtained by varying the
action (\ref{eq:IGbrane}) with respect to the background metric $\gamma_{\lambda\rho}$,
which is given by
\[
T_{\lambda\rho}=\kappa^{-1}G_{\lambda\rho},
\]
where $G_{\lambda\rho}$ is the Einstein tensor of the background.

Therefore if one attempts to turn the G-brane into a source for the
gravitational field, the Einstein's equations are identically satisfied.
Hence the coupling of the G-brane to its ancestor is not achieved.
This is quite in line with the interpretation of the G-brane as gravitation
without gravitons, because when one turns on the full dynamical gravitational
field, the G-brane should be automatically included rather than being
an additional ``matter'' source.

\section{G-brane for $D=2$\label{sec:G-brane-for}}

\subsection{Action}

As we saw in the previous section, for spacetime dimensions $D\geq3$
the G-brane action is a descendant of the action for the gravitational
field. For $D=2$ there is an analog of the Einstein theory of gravitation,
which has similarities with its higher dimensional counterpart but
also possesses key differences with it. The similarities and differences
of their gravitational ancestors are inherited by their descendants,
the G-branes.

For two spacetime dimensions the analog of the Einstein equation,
\[
R_{\mu\nu}-\frac{1}{2}g_{\mu\nu}R+\Lambda g_{\mu\nu}=0,
\]
 is, 
\begin{equation}
R-\Lambda=0.\label{eq:Rlambda}
\end{equation}
In both cases the sign of $R$ is the same as the sign of $\Lambda$.

Equation (\ref{eq:Rlambda}) may be derived \cite{Teitelboim-2D,Jackiw_2D}
from an action which is built out of the spacetime metric $g_{\mu\nu}$.
That action shares with its higher dimensional counterparts the properties
of depending only on the first derivative of the spatial metric $g_{11}$
and of containing no time derivatives of the $g_{0\mu}$; but it has
the important difference of not being invariant under spacetime reparametrizations.

However the lack of invariance is of such a especial nature that it
makes it still possible to define from it a consistent set of surface
deformation generators. The price paid is the appearance of a central
charge in the surface deformation algebra, which has the consequence
that the generators cannot be demanded to vanish. Due to the absence
of initial value constraints, $D=2$ gravity has one independent degree
of freedom per point, rather than ``minus one'' as extrapolated
count from higher dimensions would naively indicate. A similar situation
will arise for its descendant, the $\left(1+1\right)$ G-brane.

In \cite{Teitelboim-2D} the metric of the two-dimensional spacetime,
referred to the ``internal'' coordinates $t,x$, was written as
\begin{equation}
g_{\mu\nu}=e^{\varphi}\begin{bmatrix}-\left(\eta^{\perp}\right)^{2}+\left(\eta^{1}\right)^{2}\quad & \quad\eta^{1}\\
\eta^{1}\quad & \quad1
\end{bmatrix}.\label{eq:gab-1}
\end{equation}
Here the rescaled lapse $\eta^{\perp}=e^{-\varphi/2}N^{\perp}$ and
the customary shift $\eta^{1}=N^{1}$, describe a generic deformation
of a parametrized one-dimensional surface embedded in a two-dimensional
spacetime in a conformally invariant manner, through the introduction
of a Weyl invariant normal $\tilde{n}^{\lambda}$ which differs from
the unit normal $n^{\lambda}$ by a scale factor. That is one writes
\begin{align}
\dot{y}^{\lambda} & =\eta^{\perp}\tilde{n}^{\lambda}+\eta^{1}y_{\:\:,1}^{\lambda},\label{eq:ypunto-1}\\
\gamma_{\lambda\rho}\tilde{n}^{\lambda}\tilde{n}^{\rho} & =-g_{11}\equiv-g.\label{eq:normal-1}
\end{align}
Under a spacetime dependent rescaling of the metric $\gamma_{\lambda\rho}\left(y\right)\rightarrow e^{\sigma\left(y\right)}\gamma_{\lambda\rho}\left(y\right)$
the function $\varphi$ changes as $\varphi\rightarrow\varphi+\sigma$,
whereas $y^{\lambda}$, $\tilde{n}^{\lambda}$, $\eta^{\perp}$, $\eta^{1}$
remain invariant. The Weyl invariant normal $\tilde{n}^{\lambda}$
is related to the unit normal used before by
\[
\tilde{n}^{\lambda}=g^{1/2}n^{\lambda}.
\]

The Lagrangian density reads
\begin{equation}
\mathcal{L}=\frac{1}{4k}\left(2e^{\varphi}K^{2}-\frac{1}{2}\varphi'^{2}+2\varphi''+\Lambda e^{\varphi}\right),\label{eq:L2D}
\end{equation}
where the extrinsic curvature $K$ is understood to be expressed in
terms of the surface variables $y^{\lambda}$, 
\[
K=g^{-3/2}\tilde{n}_{\lambda}D_{1}y_{\;\;,1}^{\lambda}\,.
\]
Here $D_{1}=y_{\;\;,1}^{\lambda}D_{\lambda}$ is the covariant derivative
in the external space projected on the $t=\text{constant}$ surface.
The ``(1+1) gravitational constant $k$'' has the dimensions of
an inverse action.

The action is given by
\[
I_{\text{grav}}=\int dtdx\eta^{\perp}\mathcal{L}.
\]

\subsection{Integrability and closure. Classical central charge}

If one reparametrizes by
\[
\delta y^{\lambda}=\xi^{\perp}\tilde{n}^{\lambda}+\xi^{1}y_{\;\;,1,}^{\lambda}
\]
the action for a spacetime region $M$ with boundary $\partial M$
changes by

\begin{equation}
\delta I_{\text{grav}}=\int_{\partial M}\xi^{\perp}\mathcal{L}+\frac{1}{k}\int_{M}\left(\eta^{\perp}\xi^{1'''}-\xi^{\perp}\eta^{1'''}\right).\label{eq:deltaI2}
\end{equation}

If the volume term on the right-hand side of the above equation were
absent, the differential $\delta I_{\text{grav}}$ would be exact,
that is the integral $I_{\text{grav}}$ would not depend on the particular
foliation of $M$ used to calculate it. This is the case for the G-brane
discussed in the preceding section and it implies that the generators,
\[
s_{\lambda}=w_{\lambda}+g^{-1}\mathcal{L}\tilde{n}_{\lambda},
\]
commute, i.e., that their Poisson bracket vanishes identically.

However when the differential $\delta I_{\text{grav}}$ is not exact,
its second functional derivatives do not commute, and therefore the
brackets of two $s_{\lambda}$ is not zero. The lack of commutativity
of the $s_{\lambda}$ translates into a modification of the surface
deformation algebra, which may be read directly from (\ref{eq:deltaI2}).
One obtains for the projected form of the generators, 
\begin{align}
\left[s_{\perp}\left(x\right),s_{\perp}\left(x'\right)\right] & =\left(s_{1}\left(x\right)+s_{1}\left(x'\right)\right)\delta'\left(x,x'\right),\label{eq:hphp-1-1}\\
\left[s_{1}\left(x\right),s_{\perp}\left(x'\right)\right] & =\left(s_{\perp}\left(x\right)+s_{\perp}\left(x'\right)\right)\delta'\left(x,x'\right)\nonumber \\
 & -\frac{1}{k}\delta'''\left(x,x'\right),\label{eq:hphi-1-1}\\
\left[s_{1}\left(x\right),s_{1}\left(x'\right)\right] & =\left(s_{1}\left(x\right)+s_{1}\left(x'\right)\right)\delta'\left(x,x'\right).\label{eq:hihi-1-1}
\end{align}

From these equations one may work ``backwards'' to obtain the Poisson
bracket of the unprojected $s_{\lambda}$. This gives
\begin{align*}
\left[s_{\lambda}\left(x\right),s_{\rho}\left(x'\right)\right] & =\frac{1}{k}\left(a_{\lambda}\left(x\right)b_{\rho}\left(x'\right)+a_{\rho}\left(x'\right)b_{\lambda}\left(x\right)\right)\times\\
 & \;\;\times\delta'''\left(x,x'\right),
\end{align*}
where $a_{\alpha}$ and $b_{\alpha}$, which are Weyl invariant, are
given by
\[
a_{\alpha}=g^{-1}\gamma_{\alpha\beta}y_{\;\;,1}^{\beta},\quad b_{\alpha}=g^{-1}\gamma_{\alpha\beta}\tilde{n}^{\beta}.
\]
One sees that the $s_{\lambda}$ indeed do not commute.

Equations (\ref{eq:hphp-1-1})--(\ref{eq:hihi-1-1}) for the Poisson
brackets of the projected generators differ from the surface deformation
algebra (\ref{eq:hphp})--(\ref{eq:hihi}) on the following aspects:

(i) The coefficient of $\delta'$ on the right side of (\ref{eq:hphi-1-1})
differs from that on (\ref{eq:hphi}), because on account of the normalization
$h_{\perp}$ has now weight two instead of weight one as previously.

(ii) The metric $g^{ij}$ drops out from the right side of (\ref{eq:hphp-1-1}).
This is a consequence of the conformally invariant normalization (\ref{eq:normal-1})
for $\tilde{n}^{\alpha}$, and it only happens in one spatial dimension.
The whole algebra is then a true algebra, in the sense that its structure
constants are indeed constants rather than being field dependent. 

(iii) A central charge appears. Since the central charge has vanishing
Poisson brackets with everything the surface deformations are still
integrable. As shown by Eq. (\ref{eq:deltaI2}) the central charge
can be read directly from the lack of reparametrization invariance
of the action. The fact that, because the charge is central, the second
term on the right side of (\ref{eq:deltaI2}) is independent of the
field $\varphi$, is responsible for the covariance of the equation
of motion (\ref{eq:Rlambda}).

The surface deformation algebra in the form (\ref{eq:hphp-1-1})--(\ref{eq:hihi-1-1})
is the Lie algebra of the conformal group, which consists of two copies
of the Virasoro algebra,
\begin{equation}
\left[L\left(x\right),L\left(x'\right)\right]=\left(L\left(x\right)+L\left(x'\right)\right)\delta'\left(x,x'\right)-\frac{1}{2k}\delta'''\left(x,x'\right),\label{eq:Vir}
\end{equation}
 whose generators are
\[
L^{\pm}\left(x\right)=\left(1/2\right)\left(s_{\perp}\left(\pm x\right)\pm s_{1}\left(\pm x\right)\right).
\]
 With the standard convention the Virasoro central charge $c$ appearing
in (\ref{eq:Vir}) is given by
\[
c=\frac{12\pi}{k}.
\]

Since, as a consequence of the presence of the central charge, the
surface deformation generators cannot be required to vanish, the $s_{\lambda}$
are not to be regarded as generators of a gauge transformation but,
rather, they generate a global symmetry transformation. The $\left(1+1\right)$
G-brane is then a novel conformal field theory with 2 degrees of freedom
$y^{\lambda}$ per point, as it is highlighted in \cite{BP-two-dimensional}.
Moreover it will also prove useful in connection with asymptotic symmetries
of (2+1) gravity formulated in terms of the G-brane, which is the
subject of the next section.

\section{Gravitation in three spacetime dimensions as the theory of a G-brane\label{sec:Gravitation-in-three}}

The standard Einstein theory of gravitation in three-dimensional spacetime
is remarkable in the sense that it keeps the formal structure of its
higher dimensional counterparts, but it has no field degrees of freedom
associated to each point of two-dimensional space at a given time.
In spite of this lack of ``bulk degrees of freedom,'' when the cosmological
constant is negative, (2+1) gravity possesses black holes. It also
has a peculiar structure at spatial infinity which is not present
in higher dimensions, namely an infinite dimensional global symmetry.

However, in spite of its simplicity, the theory has so far only been
solved explicitly in a reformulation in terms of a Chern-Simons connection.
The Chern-Simons description has its own elegance and interest but,
when one develops it, becomes disconnected from the original metric
formulation and hence from spacetime.

The purpose of this section is to present a different reformulation
of the theory, which is very close to, and actually suggested by,
the metric formulation. Its spacetime significance is direct, even
more direct than that in terms of the metric, and it stays manifest
throughout. The reformulation also permits to solve the theory explicitly.

In three spacetime dimensions there is locally only one solution of
Einstein's equations in vacuum, namely a spacetime of constant curvature,
which up to identifications is anti--de Sitter space ($\Lambda<0$),
flat space ($\Lambda=0$), or de Sitter space ($\Lambda>0$). The
absence of bulk degrees of freedom suggests that the theory should
have a G-brane formulation in terms of slices through a spacetime
of constant curvature. We shall focus on the case $\Lambda<0$ because
it has more structure: black holes and surface dynamics at spatial
infinity.

\subsection{Surface variables as potentials that solve the constraints}

Our goal is thus to formulate the theory employing the potentials
$y^{\lambda}$ and their conjugate momenta $w_{\lambda}$ as the fundamental
variables, instead of the spatial metric $g_{ij}$ and its conjugate
momentum $\pi^{ij}$. To that end we start with the ``bulk'' gravitational
action in Hamiltonian form,

\begin{equation}
I_{\text{grav}}=\int dtdx\left(\pi^{ij}\dot{g}_{ij}-N^{\perp}\mathcal{H}_{\perp}-N^{i}\mathcal{H}_{i}\right),\label{eq:Igrav}
\end{equation}
where
\begin{align*}
\mathcal{H}_{\perp} & =2\kappa G_{ijkl}\pi^{ij}\pi^{kl}-\frac{1}{2\kappa}g^{1/2}\left(R-2\Lambda\right),\\
\mathcal{H}_{i} & =-2\pi_{i\;/j}^{\; j}\,,
\end{align*}
with
\[
G^{ijkl}G_{klrs}=\frac{1}{2}\left(\delta_{r}^{i}\delta_{s}^{j}+\delta_{s}^{i}\delta_{r}^{j}\right).
\]

If one substitutes in (\ref{eq:Igrav}) the relationship between the
momentum and the velocity,
\[
\pi^{ij}=-\frac{1}{2\kappa}G^{ijkl}K_{kl},
\]
one recovers the Lagrangian form (\ref{eq:L}).

The key observation now is that, if one takes $\gamma_{\lambda\rho}$
to be the anti--de Sitter metric, and substitutes expressions (\ref{eq:saptialmetric})
and (\ref{eq:extrinsic curv}) in the generators $\mathcal{H}_{\perp}$
and $\mathcal{H}_{i}$ appearing in (\ref{eq:Igrav}), then they vanish
identically, 
\begin{align}
\mathcal{H}_{\perp} & \equiv0,\label{eq:Hp}\\
\mathcal{H}_{i} & \equiv0.\label{eq:Hi}
\end{align}

Conversely the most general solution of the constraint equations is
obtained in this way, because they are precisely the equations of
Gauss and Codazzi, which guarantee local embeddability. One may think
of the Gauss-Codazzi equations implying the existence of the potentials
$y^{\lambda}$ for $g_{ij}$ and $\pi^{ij}$, as the analog of the
Poincaré lemma stating that if a form is closed then it is locally
exact.

Note that the $y^{\lambda}$ have more information than that contained
in the $g_{ij}$ and the $\pi^{ij}$ because, when the latter are
given, one needs to specify in addition the location of the surface
at infinity in order to determine $y^{\lambda}$ everywhere. This
means that in the G-brane formulation ``the problem of time'' namely
the reconstruction of the proper time separation between two surfaces
from the knowledge of their intrinsic and extrinsic geometry is automatically
solved \cite{wheeler}. See also \cite{MTW}.

After solving the constraints, only the $\pi^{ij}\dot{g_{ij}}$ term
is left in the bulk Hamiltonian action. Integrating by parts in space
and using (\ref{eq:Hi}) one obtains the action
\begin{equation}
I=-\frac{1}{\kappa}\int dtdx\left(G^{ijkl}K_{ij}K_{kl}n_{\lambda}\dot{y}^{\lambda}\right),\label{eq:Ipi2}
\end{equation}
which, on account of (\ref{eq:Hp}) may be rewritten as 
\begin{equation}
I=-\frac{1}{\kappa}\int dtdx\left(g^{1/2}\left(R-2\Lambda\right)n_{\lambda}\dot{y}^{\lambda}\right).\label{eq:Ir}
\end{equation}
Equations (\ref{eq:Ipi2}) and (\ref{eq:Ir}) are alternative forms
of the G-brane action (\ref{eq:IGbrane}) because now the background
is ``on shell,'' and therefore,
\[
G^{ijkl}K_{ij}K_{kl}=g^{1/2}\left(R-2\Lambda\right).
\]

Thus we have reformulated (2+1) gravity as the theory of a (2+1) G-brane.

\subsection{The (2+1) black hole as a G-brane\label{sub:The-(2+1)-black}}

To gain further insight into how the G-brane formulation captures
key features of the theory, we examine in its context the (2+1) black
hole \cite{BTZ,HBTZ}. Following \cite{Carlip-Teitelboim} we take
for the background metric the anti--de Sitter in the form
\[
ds_{AdS}^{2}=\frac{\ell^{2}}{\sin^{2}\chi}\left[d\chi^{2}+d\Phi^{2}-\cos^{2}\chi d\Theta^{2}\right].
\]
The surface variables will then be $y^{\lambda}=\left(\Theta,\Phi,\chi\right)$.

Setting
\begin{align}
\Theta & =\frac{r_{+}}{\ell^{2}}t+\frac{r_{-}}{\ell}\phi=\Theta_{0}+\Theta_{-1}\phi,\label{eq:Theta}\\
\Phi & =\frac{r_{-}}{\ell^{2}}t+\frac{r_{+}}{\ell}\phi=\Phi_{0}+\Phi_{-1}\phi,\label{eq:Phi}\\
\chi & =\arcsin\left[\left(\frac{r_{+}^{2}-r_{-}^{2}}{r^{2}-r_{-}^{2}}\right)^{1/2}\right]\label{eq:chi}
\end{align}
yields the black hole line element, 
\[
ds_{\text{black hole}}^{2}=-f^{2}dt^{2}+\frac{dr^{2}}{f^{2}}+r^{2}\left(d\phi+N^{\phi}dt\right)^{2},
\]
where
\begin{eqnarray*}
f^{2} & = & \frac{\left(r^{2}-r_{+}^{2}\right)\left(r^{2}-r_{-}^{2}\right)}{r^{2}\ell^{2}},\\
N^{\phi} & = & \frac{r_{+}r_{-}}{\ell r^{2}},
\end{eqnarray*}
provided one identifies 
\[
\phi\sim\phi+2\pi.
\]
The following observations are pertinent: 

(i) The functions $\Theta$ and $\Phi$ are not periodic in $\phi$.
They possess ``minus one modes'' of strength $r_{-}/\ell$ and $r_{+}/\ell$
respectively. This may be interpreted as the imprint of a monopole
type source. It is quite satisfactory to see the $r_{+}$ and $r_{-}$
appearing as fluxes in this manner.

(ii) If one decrees that the zero mode $\Theta_{0}$ of $\Theta$
be canonically conjugate to the minus one mode of $\Phi$, and vice
versa, that is, 
\begin{align}
\left[\Theta_{0},r_{+}\right] & =\frac{\kappa}{2\pi},\label{eq:simp1}\\
\left[\Phi_{0},r_{-}\right] & =\frac{\kappa}{2\pi},\label{eq:simp2}
\end{align}
and sets,
\begin{align*}
H & =\frac{\pi}{\kappa\ell^{2}}\left(r_{+}^{2}+r_{-}^{2}\right),\\
J & =-\frac{2\pi}{\kappa\ell}r_{+}r_{-},
\end{align*}
then one has, 
\begin{align*}
\frac{\partial\Theta}{\partial t} & =\left[\Theta,H\right],\quad\quad\frac{\partial\Theta}{\partial\phi}=-\left[\Theta,J\right],\\
\frac{\partial\Phi}{\partial t} & =\left[\Phi,H\right],\quad\quad\frac{\partial\Phi}{\partial\phi}=-\left[\Phi,J\right].
\end{align*}
This is again satisfactory since $H$ and $J$ are precisely the mass
and the angular momentum of the black hole. The symplectic structure
(\ref{eq:simp1}) and (\ref{eq:simp2}) will be useful in the more
general setting of Sec. \ref{sub:Coupling-of-the} below.

\subsection{Symmetry, degrees of freedom and dynamics at spatial infinity}

In 2+1 dimensions with a negative cosmological constant, the gravitational
field has an asymptotic symmetry at large spacelike distances whose
Lie algebra is infinite dimensional \cite{Brown-Henneaux}. This is
a peculiarity of this low dimension, because in higher dimensions
the corresponding symmetry would just be that of anti--de Sitter space.
It was shown in \cite{Brown-Henneaux} that, in (2+1) spacetime dimensions,
the boundary of an ``asymptotically anti--de Sitter space'' may
be taken to be a flat cylinder. At large spatial distances the surface
variables of the (2+1) G-brane describe asymptotically a generic cut
of that flat cylinder. This simple statement captures in a nutshell
the symmetries of the boundary. The generic cut may be implemented,
for example, by allowing the functions $\Theta$ and $\Phi$ appearing
in (\ref{eq:Theta}) and (\ref{eq:Phi}) to possess all the Fourier
modes higher than 1. The asymptotic frame in which the higher modes
are absent and (\ref{eq:Theta}) and (\ref{eq:Phi}) hold could be
called the ``rest frame.'' One may go from the rest frame to a generic
frame by means of a ``boost,'' generated by the higher Fourier modes
of the asymptotic symmetry generators. 

{[}Incidentally, the previous remarks underline the fact that the
presence of the infinite-dimensional asymptotic symmetry algebra,
and of the surface field degrees of freedom that accompany it for
$D=3$, does not account for the Bekenstein-Hawking entropy any more
than the states of different linear momentum does for $D\geq4$, in
which case the asymptotic symmetry algebra is finite dimensional (see
\cite{Regge-Teitelboim} for $\Lambda=0$, and \cite{HT-AdS} for
$\Lambda<0$). One would rather expect the leading contribution to
the entropy to come, in a way yet to be properly understood, from
the zero mode (see \cite{Teitelboim-statistical} in this context){]}. 

In order to be realized canonically, the infinite dimensional symmetry
algebra requires an infinite number of surface degrees of freedom,
which are present although the gravitational field does not have bulk
degrees of freedom. These degrees of freedom emerge naturally by recalling
that the condition for flatness of the cylinder at infinity is precisely
the equation of motion of the (1+1) gravity theory discussed in the
previous section, for $\Lambda=0$, that is,
\begin{equation}
^{\left(2\right)}R=0.\label{eq:Rcero}
\end{equation}
 Therefore the boundary degrees of freedom are contained in the field
$\varphi$ appearing in (\ref{eq:gab-1}).

If one passes from the Lagrangian density (\ref{eq:L2D}) to the Hamiltonian
one finds 
\begin{equation}
H=\int dx\left(\eta^{\perp}t_{\perp}+\eta^{1}t_{1}\right),\label{eq:HHH}
\end{equation}
with
\begin{align}
t_{\perp} & =2k\pi^{2}+\frac{1}{8k}\varphi'^{2}-\frac{1}{2k}\varphi'',\label{eq:hhperp}\\
t_{1} & =\pi\varphi'-2\pi',\label{eq:hhi}
\end{align}
which obey the same algebra (\ref{eq:hphp-1-1})-(\ref{eq:hihi-1-1}).
In the Hamiltonian formulation these generators are the ancestors
of the (1+1) G-brane generators appearing in (\ref{eq:hphp-1-1})-(\ref{eq:hihi-1-1}).
Actually, of course, the reasoning is the other way around: the algebra
of the G-brane generators $s_{\perp},s_{1}$ is inherited from that
of its ascendants $t_{\perp},t_{1}$.

\subsection{Surface integrals as symmetry generators and their relation with
the (1+1) G-brane}

In order to account for the most general permissible motion at infinity,
one must add to the bulk Hamiltonian of gravity a surface term at
infinity \cite{Regge-Teitelboim}. The surface term for (2+1) gravity
was given in \cite{Brown-Henneaux} and takes the form
\begin{equation}
H=\int_{\infty}d\phi\left(\eta^{\perp}\mathcal{H}_{\perp\infty}+\eta^{1}\mathcal{H}_{1\infty}\right).\label{eq:Hboundary}
\end{equation}

In general, the surface term, which is the generator of asymptotic
motions, may be realized canonically only if one fixes the coordinate
system in the bulk relative to the one at infinity, and employs the
Dirac bracket. It was found in \cite{Brown-Henneaux} that $\mathcal{H}_{\perp\infty},\mathcal{H}_{1\infty}$
obey the algebra (\ref{eq:hphp-1-1})--(\ref{eq:hihi-1-1}) with a
central charge, 
\begin{equation}
k=\frac{\kappa}{\ell}.\label{eq:kappa}
\end{equation}
In the present context, \ref{eq:kappa} may be thought of as determining
the (1+1) gravitational constant $k$ in terms of its (2+1) counterpart
$\kappa$.

One may realize canonically the generators of asymptotic motions precisely
by means of the generators $t_{\perp},t_{1}$, by simply \emph{defining}
the variables $\varphi$ and $\pi$ in terms of the asymptotic parts
of $\pi^{ij}$ and $g_{ij}$ appearing in $\mathcal{H}_{\perp\infty}$
and $\mathcal{H}_{1\infty}$ through 
\begin{align}
t_{\perp} & =\mathcal{H}_{\perp\infty},\nonumber \\
t_{1} & =\mathcal{H}_{1\infty,}\label{eq:hh}
\end{align}
regarded as differential equations for $\varphi$ and $\pi$ with
$\mathcal{H}_{\perp\infty}$ and $\mathcal{H}_{1\infty}$ given. 

The realization (\ref{eq:hh}) may be considered as ``intrinsic,''
or ``gauge invariant,'' in the sense that no reference is made in
it of the boundary conditions used in approaching the boundary, or
of the way in which the ``bulk slicing'' is fixed in order to calculate
a Dirac bracket. The fact that the cylinder at infinity is the boundary
of the bulk is imprinted through Eq. (\ref{eq:kappa}).

\subsection{Treatment of zero modes}

To implement (\ref{eq:hh}) it is useful to treat the zero modes separately.
That is we will assume that both $\varphi$ and $\pi$ do not possess
a zero mode when expanded in a Fourier series, and reinstate afterwards
the zero modes by replacing

\begin{align}
\varphi' & \rightarrow\varphi'-\frac{2}{\ell}r_{+},\label{eq:phizaeromodes}\\
\pi & \rightarrow\pi+\frac{1}{2k\ell}r_{-}\label{eq:pizeromodes}
\end{align}
in the Hamiltonian generators (\ref{eq:hhperp}) and (\ref{eq:hhi}).
Then in (\ref{eq:hh}) there are as many equations as unknowns. The
anti--de Sitter radius $\ell$ is brought into the above equations
because we have replaced the spatial coordinate $x$ appearing in
(\ref{eq:HHH}) by an angle $\phi$ by setting $x=\ell\phi$.

The complete action takes then the form

\begin{equation}
I=\int dt\left(\frac{2\pi}{\kappa}r_{+}\dot{\Theta}_{0}+\frac{2\pi}{\kappa}r_{-}\dot{\Phi}_{0}+\int d\phi\pi\dot{\varphi}-H\right),\label{eq:Izeromodes}
\end{equation}
where $H$ is given by (\ref{eq:Hboundary}) with the replacements
(\ref{eq:phizaeromodes}) and (\ref{eq:pizeromodes}) implemented.
When $\varphi$ and $\pi$ are set equal to zero the dynamics derived
from (\ref{eq:Izeromodes}) is just the one described in Sec. \ref{sub:The-(2+1)-black}.

In the treatment just presented, because the boundary is flat, one
is naturally led by (\ref{eq:Rcero}) to a description in terms of
a free field $\varphi$ which has a direct metric interpretation %
\footnote{According to (\ref{eq:gab-1}) the field is the logarithm of the metric
component of a cut of the boundary. To reach that cut one would have
to tune the $y^{\lambda}$ coordinate system so that, as the boundary
is approached, it matches the ``intrinsic frame'' selected by (\ref{eq:hh}).
For example, the coordinates which realize the boundary conditions
of \cite{Brown-Henneaux} will not do. That a coordinate system that
does the job indeed exists is guaranteed by the fact that the dynamics
of $\varphi$ ensures that the boundary is flat. %
}. This field may be thought of as the aggregate of the two chiral
bosons emerging at the end of the analysis of the asymptotic structure
of the Chern-Simons formulation given in \cite{Henneaux-Maoz-Schwimmer}
where, incidentally, accounting for the ``holonomies'' $r_{+},r_{-}$
remained a bit of an issue. See also \cite{Carlip} and references
therein.

Within the Chern-Simons framework, a different proposal for the dynamics
of the boundary has also been arrived at \cite{Coussaert-Henneaux-vanDriel}.
It is the Liouville theory, in which a term proportional to $e^{\varphi}$
is included in the Hamiltonian generator (\ref{eq:hhperp}). In the
present context this would correspond to replacing (\ref{eq:Rcero})
by 

\[
^{\left(2\right)}R=\text{constant}.
\]

Although we have not explored this possibility, it is tempting to
speculate that such a term could arise by replacing the flat cylinder
at infinity by one of constant curvature, through a different choice
of boundary conditions. This would be in line with the fact that one
can canonically relate the Liouville theory with the one corresponding
to $^{\left(2\right)}R=0$ by a Bäcklund transformation (see e.g \cite{Jackiw_2D}
and references therein).

\subsection{Coupling of the boundary degrees of freedom to a boundary G-brane\label{sub:Coupling-of-the}}

The description of the boundary dynamics and the asymptotic symmetries
in terms of a (2+1) G-brane becomes sharper if one introduces a (1+1)
G-brane on the boundary. The key point is the observation that the
combinations,
\begin{align}
h_{\perp} & =\left.s_{\perp}\right|_{k\rightarrow-k}+t_{\perp},\label{eq:hhperp-1}\\
h_{1} & =s_{1}+t_{1},\label{eq:hh11}
\end{align}
close according to the algebra (\ref{eq:hphp-1-1})--(\ref{eq:hihi-1-1})
\emph{without central charge}. Here the notation $k\rightarrow-k$
means that one replaces in $s_{\perp}$ the constant $k$ by its negative.
This replacement maintains the surface deformation algebra, but changes
the sign of the central charge.

Since the algebra of the generators $h_{\perp},h_{1}$ has no central
charge, it is consistent to demand that these $h$'s be constrained
to vanish,
\begin{align}
h_{\perp} & \approx0,\label{eq:cp}\\
h_{1} & \approx0.\label{eq:c1}
\end{align}
Thus the canonically conjugate field $\varphi$ and $\pi$ out of
which the generators $t_{\perp}$ and $t_{1}$ are constructed may
be regarded as matter fields defined on the (1+1) G-brane and accompanying
it through its deformations. 

In terms of the above generators, the action for (2+1) gravity becomes
the sum of three terms, integrated respectively over manifolds of
dimension three (the bulk), two (the boundary), and unity (the history
of the zero modes): 
\begin{align}
I & =\int dt\left[\int d^{2}x\: w_{\lambda}\dot{y}^{\lambda}+\int d^{1}x\:\left(w_{a}\dot{y}^{a}+\pi\dot{\varphi}\right)\right.\nonumber \\
 & \quad\left.+\frac{2\pi}{\kappa}r_{+}\dot{\Theta}+\frac{2\pi}{\kappa}r_{-}\dot{\Phi}-H\right],\label{eq:Imatching}
\end{align}
with
\begin{align*}
H & =\int d^{2}x\left(N^{\perp}h_{\perp}^{\left(2+1\right)}+N^{i}h_{i}^{\left(2+1\right)}\right)\\
 & \quad+\int dx\left(\eta^{\perp}h_{\perp}^{\left(1+1\right)}+\eta^{1}h_{1}^{\left(1+1\right)}\right),
\end{align*}
and
\begin{align*}
\eta^{\perp} & =g^{-1/2}N^{\perp}\left(\infty\right),\\
\eta^{1} & =N^{1}\left(\infty\right).
\end{align*}

In (\ref{eq:Imatching}) we have allowed for a generic background
coordinate system $y^{a},\;\left(a=0,1\right)$, with corresponding
metric $\gamma_{ab}$, on the flat boundary at infinity. If the boundary
were described as $y^{\lambda}=y^{\lambda}\left(t,\phi,r_{0}\right)$
with $r_{0}\rightarrow\infty$, then the $y^{a}$ would be related
to $t$ and $\phi$ by a coordinate transformation, which might depend
on $r_{0}$ as a parameter.

The formulation of (2+1) gravity in terms of (2+1) and (1+1) G-branes
summarized in the action (\ref{eq:Imatching}) is ``fully parametrized''
in the sense of possessing a Hamiltonian that vanishes weakly, even
when the boundary dynamics is included {[}the boundary action is invariant
under spacetime reparametrizations because due to the change $k\rightarrow-k$
in (\ref{eq:hhperp-1}), the terms of the form (\ref{eq:deltaI2})
coming from the brane and the matter contributions mutually cancel{]}.

Such a description would not have been possible if one had attempted
to achieve it through a Dirac or a Nambu brane, because then the central
charge would not drop out. In other words, the ``background energy''
of the (1+1) G-brane is essential. Note that the background energy
which actually appears in this construction is $-\mathcal{L}$, the
\emph{negative} of the standard one, because the sign of $k$ in $s_{\perp}$
has been changed. There is no harm in this because $\mathcal{L}$
does not have a definite sign to begin with.

\section{Concluding remarks\label{sec:Concluding-remarks}}

We have introduced, for any spacetime dimension $D$, the concept
of a G-brane, which is a space-filling brane whose action is a direct
descendant of the action for the gravitational field. The G-brane
may be thought of as the remanent of the gravitational field when
the propagating gravitons are removed. We were led to the G-brane
in the midst of an effort to gain understanding of the puzzling feature,
evidenced by the existence of a black hole in 2+1 spacetime dimensions
\cite{BTZ,HBTZ}, that key aspects of gravitation are retained even
when there are no propagating gravitational field degrees of freedom.
The G-brane concept was applied for the lower dimensions $D=3$ and
$D=2$, where the following results were obtained: (i) For $D=3$
the G-brane yields a reformulation of gravitation theory in which
the Hamiltonian constraints can be solved explicitly, while keeping
the spacetime structure manifest. (ii) For $D=2$ one finds a new
realization of the conformal algebra, i.e. a conformal field theory,
in terms of two scalar fields and their conjugates, which possesses
a classical central charge. (iii) In the G-brane reformulation of
(2+1) gravity, the boundary degrees of freedom of the gravitational
field in asymptotically anti--de Sitter space appear as matter coupled
to the (1+1) G-brane on the boundary.

Finally, a word about future developments. As it was just indicated,
although the concept and general properties of the G-brane were presented,
no applications were given for the higher dimensions $D\geq4$. Therefore
a problem that brings itself immediately for analysis is how, for
$D\geq4$, the G-brane inherits from the gravitational field two key
aspects, namely, the entropy and the asymptotic structure at spacelike
infinity. Additionally one would like to develop the notion of a supersymmetric
G-brane for all $D\geq2$. Finally a less well-defined question is
to ask if and how, the idea of the G-brane as ``gravitation without
gravitons'' relates to two recent different lines of inquiry, namely:
(i) the role of ``soft gravitons'' and their relation with asymptotic
symmetries discussed in \cite{Strominger1} and \cite{Strominger2}
and, (ii) the interplay of ``longitudinal gravitons'' with black
hole entropy discussed in \cite{Dvali1} and \cite{Dvali2}. We hope
to address these issues in the near future. 
\begin{acknowledgments}
The Centro de Estudios Científicos (CECs) is funded by the Chilean
Government through the Centers of Excellence Base Financing Program
of Conicyt. C.B. wishes to thank the Alexander von Humboldt Foundation
for a Humboldt Research Award. The work of A.P. is partially funded
by the Fondecyt Grant Nº 11130262. \end{acknowledgments}

\end{document}